\documentclass[aps,a4paper,showkeys,nofootinbib,longbibliography,notitlepage,superscriptaddress,twocolumn]{revtex4-2}
\usepackage{graphicx} 
\usepackage{dcolumn} 
\usepackage{bm} 
\usepackage{braket} 
\usepackage[ISO]{diffcoeff}
\usepackage[usenames,dvipsnames]{color}
\usepackage[
colorlinks=true, citecolor=blue, urlcolor=blue, linkcolor=blue,
setpagesize=false, bookmarks=false
]{hyperref}
\usepackage{siunitx}
\usepackage{mhchem}
\usepackage{mathtools}
\usepackage{overpic}

\usepackage[dvipsnames]{xcolor}
\usepackage{hyperref} 
\usepackage{cleveref} 
\newcommand{\sectionprl}[1]{{\par\it #1.---}}

\begin{document}

\title{Prethermal inverse Mpemba effect}
\author{Koudai Sugimoto}
\email{sugimoto@rk.phys.keio.ac.jp}
\affiliation{Department of Physics, Keio University, Yokohama, Kanagawa 223-8522, Japan}

\author{Tomotaka Kuwahara}
\email{tomotaka.kuwahara@riken.jp}
\affiliation{Analytical Quantum Complexity RIKEN Hakubi Research Team, RIKEN Center for Quantum Computing (RQC), Wako, Saitama 351-0198, Japan}
\affiliation{PRESTO, Japan Science and Technology (JST), Kawaguchi, Saitama 332-0012, Japan}
\author{Keiji Saito}
\email{keiji.saitoh@scphys.kyoto-u.ac.jp}
\affiliation{Department of Physics, Kyoto University, Kyoto 606-8502, Japan}


\begin{abstract}
The inverse Mpemba effect is a counterintuitive phenomenon in which a system, initially in thermal equilibrium and prepared at different temperatures below that of the final equilibrium state, relaxes to the final state more rapidly when starting from a lower initial temperature.
We extend this concept to the relaxation toward a prethermal state in isolated quantum systems.
By examining a simple model that exhibits prethermalization, we demonstrate that this effect indeed manifests under periodic driving. We further discuss the realization of this phenomenon in a variety of systems within a unified theoretical framework.
\end{abstract}

\maketitle

\sectionprl{Introduction}
The problem of thermalization in statistical mechanics remains a deep and challenging issue in modern physics.
Especially, understanding how an isolated or open system relaxes from a far-from-equilibrium initial state toward thermal equilibrium is nontrivial, and continues to motivate both theoretical and experimental advances \cite{Nandkishore2015ARCMP, Mori2018JPB}.

\begin{figure}[b]
  \centering
  \includegraphics[width=\columnwidth]{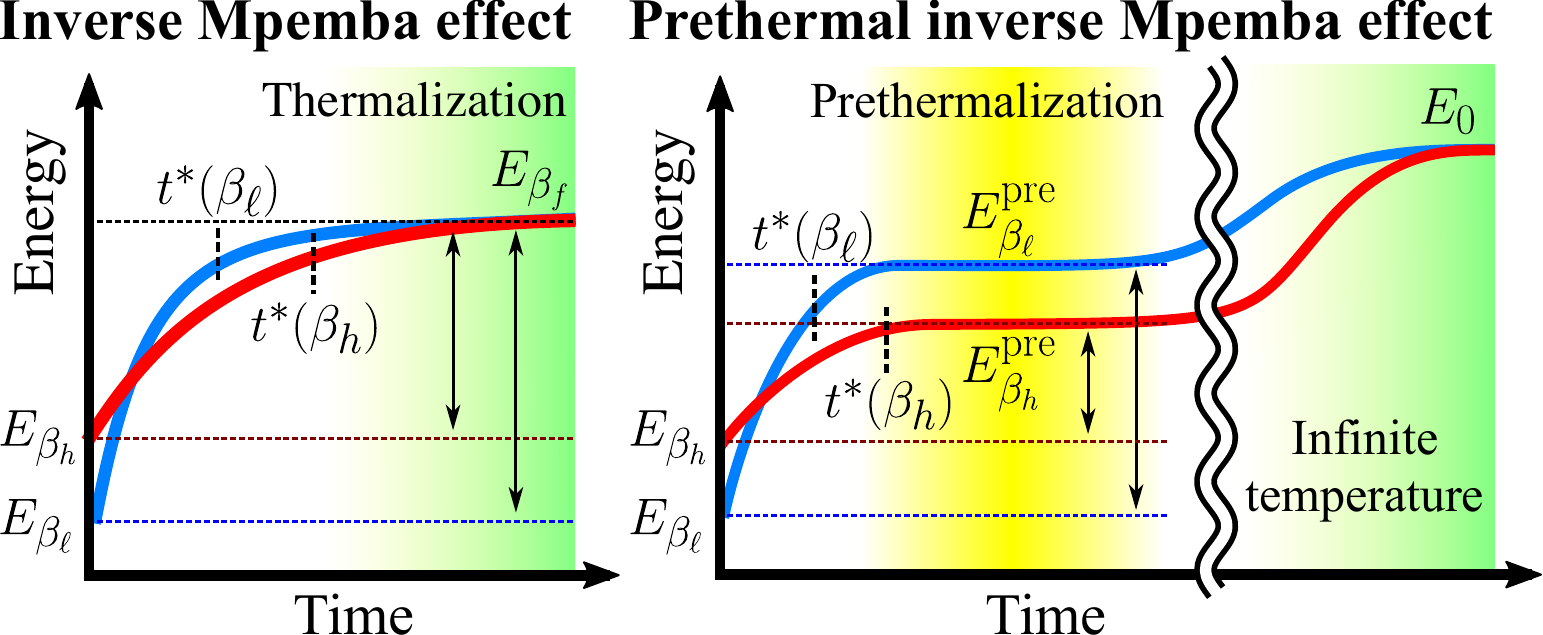}
  \caption{
  Schematic pictures of conventional IME (left panel) and IME induced by prethermalization (right panel).
  If the time evolution leads to the prethermal state and the relationship between the initial and prethermal state energies is reversed for different initial temperatures, the crossing of the energy curve is observed.
  }
  \label{fig:intro}
\end{figure}

In recent years, growing attention has been directed toward counterintuitive phenomena known as the Mpemba effect and the inverse Mpemba effect (IME) \cite{Mpemba1969PE, Lu2017PNAS, Kumar2020Nature, Teza2025arXiv, Ares2025arXiv}.
In its standard form, it refers to the situation where a system initially cools
faster at a higher temperature than an identical system starting at a lower temperature. Conversely, the IME means faster heating from a colder state (the left panel in Fig.~\ref{fig:intro}). 
These effects highlight that the relaxation dynamics can strongly depend on the initial condition, and that the timescale of thermalization is not necessarily monotonic in the initial temperature.

Although the underlying mechanism may seem simple, uncovering when and why such effects emerge in specific physical systems provides valuable insight into the complex nature of thermalization.
Recent studies have explored the Mpemba effect in a wide variety of physical setups, with growing attention devoted to quantum systems in particular.
The quantum Mpemba effect has been theoretically predicted in systems including quantum dots \cite{Chatterjee2023PRL, Wang2024PRR, Graf2025JPCM, Strachan2024arXiv}, multi-level systems \cite{Carollo2021PRL, Manikandan2021PRR, Ivander2023PRE, Chatterjee2024PRA}, spin or bosonic systems \cite{Nava2019PRB, Caceffo2024JSM, Moroder2024PRL, Bhore2025arXiv, Westhoff2025arXiv}, random quantum circuits \cite{Liu2024PRL, Turkeshi2024arXiv}, and so on \cite{chetrite2021metastable,Holtzman2022,PhysRevA.111.022215,e27040407}, and experimentally observed in platforms of single trapped ion \cite{AharonyShapira2024PRL, Zhang2025NC} and trapped-ion quantum simulators \cite{Joshi2024PRL}.
While many of these works focus on open quantum systems described by Liouvillian dynamics, where the system is in contact with a thermal bath, it is equally important to investigate the relaxation behavior in isolated quantum systems to fully understand the genuine quantum features of the Mpemba effect.
In this direction, the entanglement asymmetry has recently been proposed as a novel manifestation of the quantum Mpemba effect \cite{Ares2023NC, Ares2023SP, Murciano2024JSM, Rylands2024PRL, Yamashika2024PRB, Yamashika2024PRA, Ares2025PRB}, sparking increasing interest in the role of coherence and entanglement in isolated quantum thermalization \cite{Liu2024arXiv, Banerjee2024arXiv, Kusuki2025JHEP, Yu2025arXiv}.

In this work, we aim to broaden the scope of the quantum Mpemba effect by proposing a new type of IME that emerges in the context of multistage relaxation in isolated quantum systems.
Specifically, we demonstrate that, as shown schematically in the right panel in Fig.~\ref{fig:intro}, while the ultimate relaxation to the true steady state may occur faster when starting from a higher (or lower) temperature, an IME can arise during the earlier stage of relaxation, namely towards a prethermal state.
This perspective suggests that the Mpemba effect may be more ubiquitous than previously thought.
In particular, such behavior can arise in broader relaxation processes, including those towards metastable states or local minima, beyond relaxation to thermal equilibrium.
Building on this perspective, the extension of the Mpemba effect to prethermalization offers renewed insights that may prove essential for leveraging such effects in quantum technologies and quantum control applications.

To concretely demonstrate our proposal, we study a one-dimensional spinless fermion model subjected to a time-periodic external field.
In this setup, a system initialized at a low temperature heats up more rapidly than one prepared at a higher temperature, thereby exhibiting an IME within the prethermal regime.
Heating phenomena in periodically driven systems are well-established~\cite{PhysRevX.4.041048,PhysRevLett.116.125301,KUWAHARA201696,PhysRevLett.116.120401}, with various experimental realizations such as cold-atom~\cite{PhysRevX.10.021044,PhysRevX.11.011057} and NMR~\cite{Peng2021,PhysRevLett.127.170603} systems.
This suggests that our model is not only theoretically sound but also experimentally accessible.

Alongside numerical simulations, we introduce a simple and universal formalism that captures the observed IME in this prethermal scenario.
Our analysis suggests that such prethermal IMEs are controllable features of nonintegrable, periodically driven quantum dynamics.
This work advances the frontier of quantum Mpemba research by introducing a new classification of the effect and sheds light on the interplay between nonequilibrium dynamics and thermalization in isolated quantum systems.

\sectionprl{Extension of the concept: Prethermal inverse Mpemba effect}
In this work, we discuss the Mpemba-like effect in the relaxation dynamics toward the prethermal state, a metastable state emerging before full thermalization.

To introduce this concept, let us first discuss the conventional Mpemba effect. The initial state is assumed to be an equilibrium distribution at inverse temperature $\beta$. In the Mpemba effect, one considers the relaxation of the system toward an equilibrium distribution corresponding to a final inverse temperature $\beta_{f}$, with associated mean energy $E_{\beta_f}$.
The relaxation time may be defined in terms of the system's convergence toward a specified reference temperature or energy.
For instance, if the energy at time $t$ is denoted by $E_{\beta}(t)$ and relaxes exponentially in time toward the final state, it is natural to define the relaxation time $t^{\ast}(\beta)$ via the condition $|E_{\beta_f} - E_{\beta} (t^{\ast})| = \delta |E_{\beta_f} - E_{\beta} (0)|$, where $\delta$ is a small positive parameter.
While the Mpemba effect usually means the cooling process, in this paper, we focus on a heating process, i.e., IME. The criterion of the conventional IME is formulated as
\begin{align}
\begin{split}
  &t^{\ast} (\beta_{\ell} )  < t^{\ast} (\beta_{h} )  \, , \\
  &E_{\beta_f} > E_{\beta_h} (0) >  E_{\beta_{\ell}} (0)  \, ,  
\end{split}
~~~{\text{(conventional IME)}}
\label{stiM}
\end{align}
where $\beta_{\ell} > \beta_{h} > \beta_f$ is assumed (the left panel in Fig.~\ref{fig:intro}). That is, a system prepared at a lower initial temperature exhibits faster heating compared to one prepared at a higher temperature.

We next consider the possible extension of this concept to the relaxation to the prethermal state.
They often arise due to the presence of impurities, emerge from integrable systems perturbed by nonintegrable interactions in isolated settings, or be induced by high-frequency periodic driving. Such states are characterized by their remarkable longevity, persisting for extended periods as a result of the inhibition of direct relaxation pathways to the true thermalized state. 
Here, we focus on the relaxation toward prethermal states that appear en route to the final infinite-temperature state in quantum isolated systems, as will be exemplified later. We shall employ the term {\it prethermal state} to denote a metastable state that precedes the attainment of the system's ultimate thermal equilibrium. Motivated by the standard formulation of the IME (\ref{stiM}), let us consider an analogous scenario in which the system, initially prepared at inverse temperature $\beta$, relaxes toward a prethermal state. The prethermal state may, in general, depend on the initial temperature.
Here, we denote the energy of the prethermal state corresponding to the initial inverse temperature $\beta$ by $E_{\beta}^{\rm pre}$.
Assuming an exponential relaxation toward the prethermal state, the relaxation time $t^{\ast}$ may be defined via the condition
$
  | E_{\beta}^{\rm pre} - E_{\beta} (t^{\ast})|
    = \delta | E_{\beta}^{\rm pre} - E_{\beta} (0) |
$.
In this paper, we use $\delta = 1/e$ for the small positive parameter.

Then, we extend the IME to the prethermal state with the condition:
\begin{align}
\begin{split}
  t^{\ast} (\beta_{\ell} )  &< t^{\ast} (\beta_{h}) \, , \\
  E_{\beta_{\ell}}^{\rm pre} - E_{\beta_{\ell}} (0) &> 
  E_{\beta_{h}}^{\rm pre} - E_{\beta_{h}} (0)  \, . 
\end{split}
~{\text{(prethermal IME)}}
\label{priM}
\end{align}
A schematic picture is illustrated in the right panel of Fig.~\ref{fig:intro}.
That is, between two systems initialized at different temperatures, the one starting from a lower initial temperature may approach the higher-energy prethermal state in a shorter time. We refer to this anomalous heating as the prethermal IME.

\begin{figure}
  \centering
  \begin{tabular}{c}
    \begin{minipage}[t]{\columnwidth}
      \includegraphics[width=0.8\columnwidth]{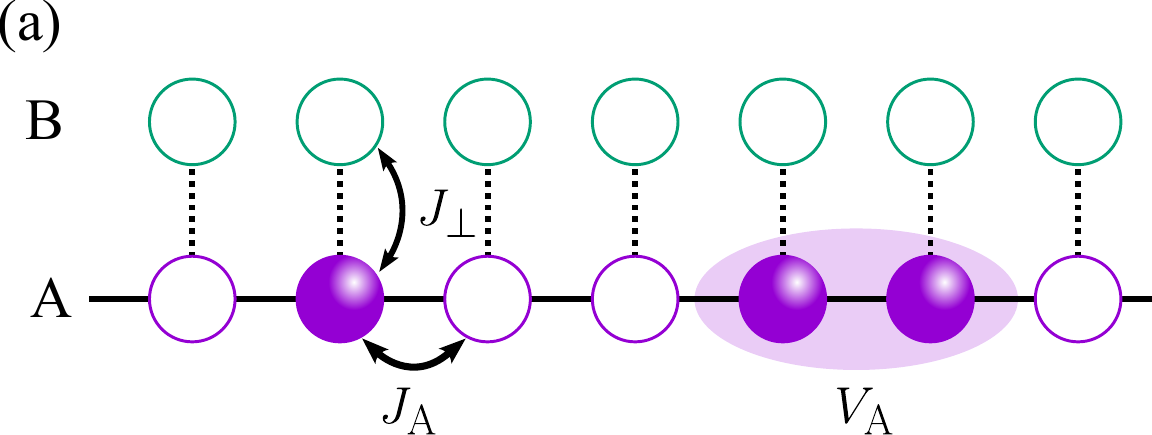}
    \end{minipage}
    \\
    \begin{minipage}[t]{\columnwidth}
      \includegraphics[width=\columnwidth]{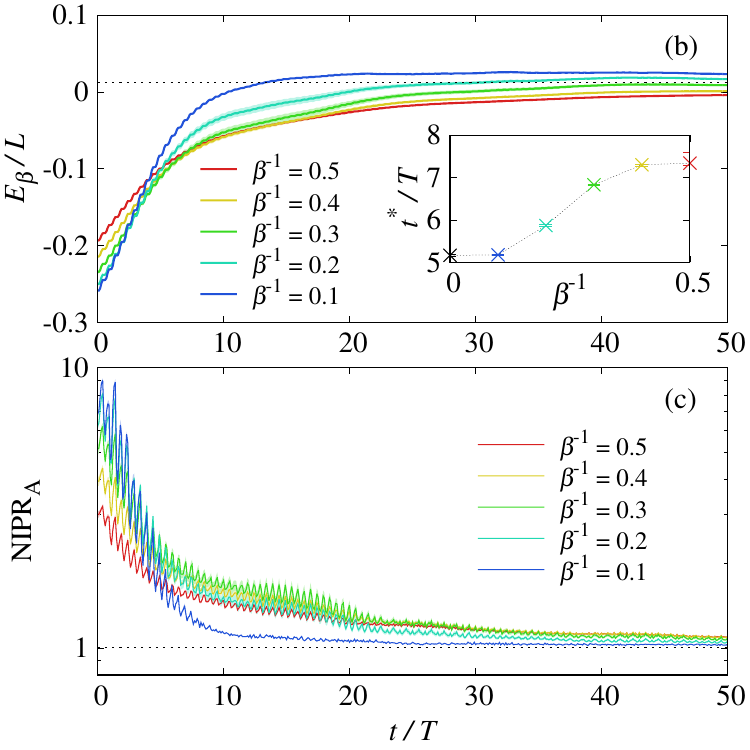}
    \end{minipage}
  \end{tabular}
  \caption{
  (a) Schematic picture of the model employed in this study.
  The total number of lattice sites and fermions is fixed to $L = 2\times 18 = 36$ and $N = 6$, respectively.
  The inset illustrates the prethermal relaxation time.
  (b) Time evolution of the energy per site for various initial temperatures.
  The black dotted line represents the energy at infinite temperature.
  (c) Time evolution of $\mathrm{NIPR}_\mathrm{A}$ for various initial temperatures.
  We employ a sample size of 10 for the ensemble average and estimate the associated standard errors using the bootstrap resampling technique.
  The resulting standard errors are indicated by the shaded areas.
  }
  \label{fig:dynamics_normal}
\end{figure}

\sectionprl{Demonstration with a simple model}
We demonstrate that the prethermal IME can occur in a simple model consisting of an isolated one-dimensional chain of spinless fermions, where each site is coupled to an ancilla site, as illustrated in Fig.~\ref{fig:dynamics_normal}(a).
We refer to the primary fermionic chain as chain A, and the set of ancilla sites as chain B.
The total Hamiltonian of the system is given by
\begin{align}
  \hat{H}_0
    &= -J_{\text{A}} \sum_{j} \left(
        \hat{c}^\dagger_{j,\text{A}} \hat{c}_{j+1,\text{A}} + \text{h.c.}
      \right)
      + V_{\mathrm{A}} \sum_{j} \hat{n}_{j,\text{A}} \hat{n}_{j+1,\text{A}} \nonumber \\
&
      -J_{\perp} \sum_{j} \left(
        \hat{c}^\dagger_{j,\text{A}} \hat{c}_{j,\text{B}} + \text{h.c.}
      \right),
\label{eq:Hamiltonian}
\end{align}
where $\hat{c}^{(\dagger)}_{j,\alpha}$ is the annihilation (creation) operator of a spinless fermion at site $j$ on chain $\alpha$ ($=\text{A}, \text{B}$), and $\hat{n}_{j,\alpha} = \hat{c}^{\dagger}_{j,\alpha} \hat{c}_{j,\alpha}$ is the corresponding number operator. The chain A incorporates nearest-neighbor interactions with coupling strength $V_{\text{A}}$.

To investigate the heating process, we apply a time-periodic spatially uniform gauge field $A(t) = A_0 \sin \Omega t$ as an external perturbation. The gauge field is introduced via the Peierls phase substitution, i.e., $J_{\text{A}} \hat{c}^\dagger_{j, \text{A}} \hat{c}_{j+1, \text{A}} \to J_{\text{A}} e^{-iA(t)} \hat{c}^\dagger_{j, \text{A}} \hat{c}_{j+1, \text{A}}$, which renders the Hamiltonian time-dependent, denoted by $\hat{H}(t)$.
To induce a prethermal IME, it is essential to engineer a prethermal state that inhibits heating. To this end, we consider the regime characterized by $J_{\perp} \ll J_{\rm A}, V_{\rm A}$, so that the relaxation between the chains A and B requires a much longer time compared to the relaxation inside the chain A. 
Since periodic driving eliminates all conservation laws, the ultimate steady state should be a high-temperature random state. However, the present 
parameter setting impedes relaxation to this final state. As a result, the system settles into a long-lived prethermal state that precedes thermalization.
In the numerical simulations, we set the parameters to $J_{\text{A}} = V_{\text{A}} = 1$ and $J_{\perp} = 0.04$.
The system size and particle number are chosen as $L = 2 \times 18 = 36$ and $N = 6$, respectively, with periodic boundary conditions imposed.
We also set $\Omega=1$ and $A_0 = 1$. 

To efficiently compute the time evolution of the internal energy in this system for finite initial temperature, 
we employ the technique using the canonical thermal pure quantum state $\ket{\psi_{\beta}} = e^{-\beta \hat{H}/2}\ket{\psi_0}$, where $\ket{\psi_0}$ is a random vector in the Hilbert space of the system \cite{Hams2000PRE, Sugiura2013PRL, Heitmann2020ZNA, Jin2021JPSJ, Iwaki2024PRB}.
The time evolution is given by
$
  \ket{\psi_{\beta} (t)}
    = \hat{\mathcal{T}} e^{-i \int^{t}_0 \hat{H} (s) \dl{s}} \ket{\psi_{\beta}}
$
with the time-ordering operator $\hat{\mathcal{T}}$, and 
the expectation value of a physical observable $\hat{O}$ at time $t$ is given as \cite{Heitmann2020ZNA, Jin2021JPSJ}
\begin{equation}
  \Braket{\hat{O}}_{\beta,t}
    = \frac
      {\overline{\braket{\psi_{\beta}(t)|\hat{O}|\psi_{\beta}(t)}}}
      {\overline{\braket{\psi_{\beta}(t)|\psi_{\beta}(t)}}},
\label{eq:evO}
\end{equation}
where the overline indicates an ensemble average over different realizations of the random vector $\ket{\psi_0}$.
The internal energy during the time evolution is defined as $E_{\beta} (t) = \braket{\hat{H}_0}_{\beta, t}$.

Figure \ref{fig:dynamics_normal}(b) shows the energy per site as a function of time with the unit of the driving period $T = 2\pi/\Omega$, for various initial temperatures. We observe the long-lived flat energy regime for many initial states. For the fixed particle number $N$, the high-temperature random state gives the energy ${V_{\text{A}}\over 2} \frac{N (N - 1)}{L (L - 1)}$ for the whole system. We draw this value with the black dotted line in Fig.~\ref{fig:dynamics_normal}(b), which clearly differs from the energies observed in the figure. Notably, for the lowest temperature case $\beta^{-1} = 0.1$, the energy exceeds this high-temperature limit value.

The inset of Fig.~\ref{fig:dynamics_normal}(b) shows relaxation times for different initial temperatures, where the prethermal energy $E^{\rm pre}_{\beta}$ is defined as the averaged energy between $t/T = 40$ and $t/T = 50$.
The occurrence of the prethermal IME is evident, for instance, by comparing the cases of $\beta^{-1}=0.1$ and $0.5$.
The relaxation to the prethermal state for $\beta^{-1}=0.1$ spans a wider energy range than that for $\beta^{-1}=0.5$, satisfying the second-line criterion in Eq.~\eqref{priM}.
In addition, the relaxation time for $\beta^{-1}=0.1$ is shorter than that for $\beta^{-1}=0.5$, meeting the first-line criterion in Eq.~\eqref{priM}.
Estimating the prethermal energy over different time intervals does not significantly affect the behavior of the prethermal relaxation time.
A crossing of energy trajectories is one of the indicators of this phenomenon.

\sectionprl{Property of the prethermal state}
We here investigate the property of the prethermal state observed in Fig.~\ref{fig:dynamics_normal}(b).
The periodic driving eventually leads to the infinite-temperature state for the entire system. As we observed numerically, the observed energy clearly differs from the high-temperature state for the entire system. Note that the metastability arises due to the weak coupling between chains A and B. Hence, we expect that the prethermal state is an almost decoupled state between each chain, with a particle number given by the initial state. Because the periodic perturbation is applied to chain A only, it is natural to expect that chain A effectively thermalizes to infinite temperature with a given initial state, while chain B does not.

We scrutinize this property using a refined diagnostic tool known as the inverse participation ratio (IPR), which quantifies the degree of randomness in a quantum state.
Given a normalized quantum state 
$\ket{\psi} = \sum_{\nu=1}^D c_{\nu} \ket{\nu}$, where $\ket{\nu} = \ket{n_1, n_2, ..., n_L}$ denotes a number basis in the Fock space of dimension $D$, the IPR is defined as 
${\rm IPR} [\psi] = \sum_{\nu} |c_{\nu} (t) |^4$.
For instance, $\mathrm{IPR} = 1$ corresponds to complete localization on a single basis state \cite{Kramer1993RPP}.
In contrast, for a completely random state, the IPR satisfies $\mathrm{IPR}[\psi] = 2/(D+1)$ \cite{Kaplan2007JPA}.
This motivates the definition of the normalized IPR (NIPR) as
$
  \mathrm{NIPR} [\psi] = \frac{D+1}{2} \, \mathrm{IPR} [\psi]
$,
such that $\mathrm{NIPR} = 1$ indicates that $\ket{\psi}$ is a fully randomized (infinite-temperature) state.
Importantly, this notion can be extended to characterize thermalization within a subsystem.
Consider the Schmidt decomposition of the quantum state
$
  \ket{\psi}
    = \sum_n \sum_{s_n}
      \lambda^{(n)}_{s_n}
        \ket{\phi^{(n)}_{s_n}}_{\text{A}}
        \ket{\varphi^{(n)}_{s_n}}_{\text{B}}
$,
where $n$ labels the particle number in chain A.
By interpreting the singular values $\lambda^{(n)}_{s_n}$ as weights of the corresponding particle-number sectors, we define the NIPR for chain A as
\begin{align}
  \mathrm{NIPR}_{\text{A}} [\psi]
    &= \sum_{n}
      \sum_{s_{n}}
      \left(\lambda^{(n)}_{s_{n}} \right)^2
      \mathrm{NIPR}[\phi^{(n)}_{s_n}].
\end{align}
See the End Matter for further details.

We plot $\mathrm{NIPR}_\mathrm{A}$ in Fig.~\ref{fig:dynamics_normal}(c).
For an initial temperature $\beta^{-1} = 0.1$, $\mathrm{NIPR}_{\text{A}}$ rapidly decreases under the time-periodic field, and reaches $\mathrm{NIPR}_{\text{A}} \approx 1$ around $t \approx 10T$, indicating that chain A has attained a fully randomized state at this time. 
Interestingly, $\mathrm{NIPR}_{\text{A}}$ also exhibits a crossing behavior analogous to that observed in the energy evolution.

\begin{figure}
  \centering
  \includegraphics[width=\columnwidth]{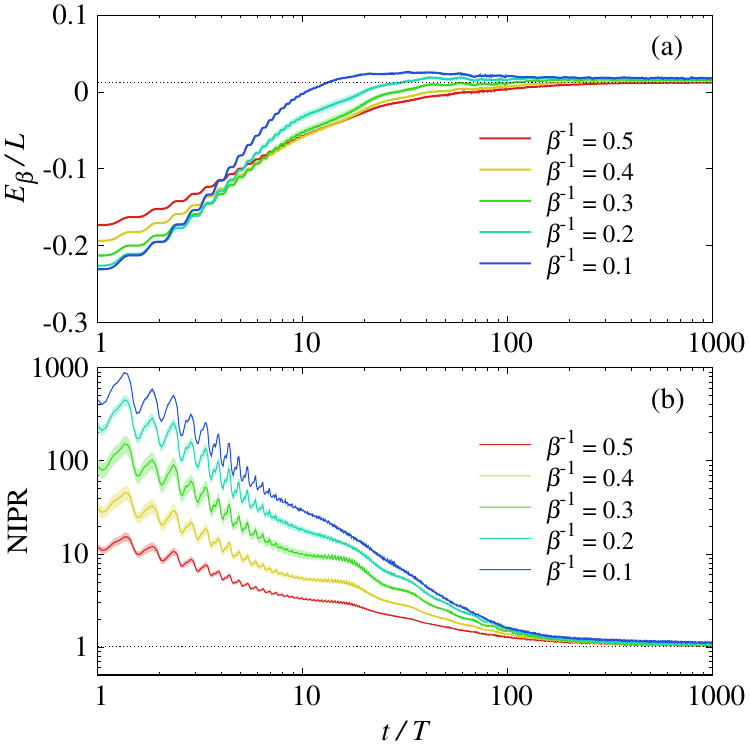}
  \caption{
  Time evolution of (a) the energy per site and (b) the NIPR for various initial temperatures.
  }
  \label{fig:dynamics_log}
\end{figure}

\sectionprl{Full thermalization in the long-time limit}
The interchain hopping $J_{\perp}$ in the present model determines the time scale of the metastability.
As $J_{\perp}$ decreases, the relaxation time to the final thermalized state of the entire system increases.
The present set of parameters enables us to see the global relaxation to the final state within the numerically accessible time window.
Figure~\ref{fig:dynamics_log}(a) shows the long-time evolution of the energy, with the $x$-axis plotted on a logarithmic scale.
For the state with an initial temperature $\beta^{-1} = 0.1$, the prethermal-state energy appears around at $t = \mathcal{O}(10T)$ and subsequently decreases, asymptotically approaching the infinite-temperature energy of the entire system, $\frac{1}{2} V_{\text{A}} \frac{N (N - 1)}{L(L - 1)}$. This observation highlights the separation of time scales: the prethermalization of chain A occurs on a much shorter time scale than the eventual global thermalization mediated by interchain coupling.
To quantify this delayed thermalization, Fig.~\ref{fig:dynamics_log}(b) shows the time evolution of the global NIPR, which gradually approaches $\mathrm{NIPR} \approx 1$ for $t \gg 100T$.
The significant difference between the time scales at which $\mathrm{NIPR}$ and $\mathrm{NIPR}_{\text{A}}$ approach 1 shows a two-stage thermalization process---from a prethermalized to a fully thermalized state.
These two-step relaxation dynamics offer compelling support for the occurrence of the prethermal IME in periodically driven quantum systems.

\sectionprl{Mechanism and generalization}
Here, we provide an intuitive explanation for this effect based on a simple toy model and discuss how the essential ingredients generalize the prethermal IME to broader physical contexts. The model must realize a phenomenon in which a system initially prepared in a lower-energy (colder) state heats up both faster and to a higher prethermal energy than one prepared in a higher-energy (hotter) state, before eventually reaching thermal equilibrium.

Consider a system composed of two weakly coupled subsystems, labeled as chain~A and chain~B. The total number of particles $N$ is conserved, so the Hilbert space is divided into sectors labeled by the particle numbers in each chain:
\begin{align}
{\cal H} = \bigoplus_{N_{\mathrm{A}} + N_{\mathrm{B}} = N} {\cal H}^{(N_{\mathrm{A}} , N_{\mathrm{B}})} \,.
\end{align}
This structure means that, during the dynamics, states are grouped according to how particles are distributed between the two chains, and transitions between sectors are strongly suppressed due to weak coupling.
Let $\gamma$ be the characteristic matrix element that couples different
$(N_{\mathrm{A}}, N_{\mathrm{B}})$ sectors.  Throughout the explanations, we consider observation times
$
  t \ll 1/\gamma ,
$
so that inter-sector transitions are rare, and each sector
effectively evolves in isolation.
Then, the numbers of particles in A and B, $N_{\mathrm{A}}$ and $N_{\mathrm{B}}$, remain nearly constant over long (but finite) timescales.  

\begin{figure}
\centering
\includegraphics[width=\columnwidth]{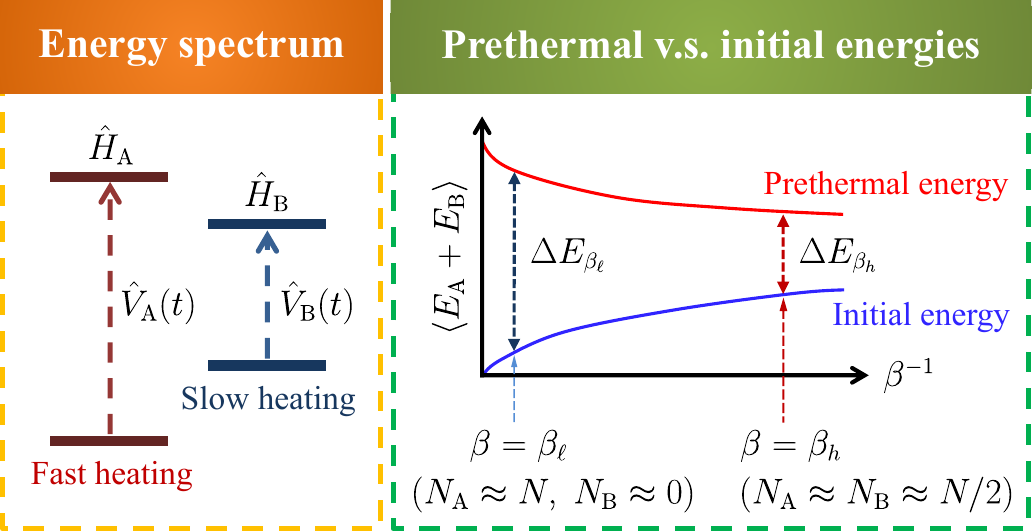}
\caption{
Schematic picture of (a) the energy spectrum ($E_{\mathrm{A}}^{\min} \preceq \hat{H}_{\mathrm{A}} \preceq E_{\mathrm{A}}^{\max}$, $E_{\mathrm{B}}^{\min} \preceq \hat{H}_{\mathrm{B}} \preceq E_{\mathrm{B}}^{\max}$; vertical axis: energy), and (b) $\beta$-dependence of the energy difference between the initial and the prethermal energies after a long (but finite) time. 
The key ingredients for the prethermal IME are: i) A conserved quantity divides the Hilbert space into weakly coupled sectors.
ii) Subsystems have overlapping but distinct energy spectra.
iii) Periodic drives are tuned so that one subsystem heats rapidly, and the other slowly. 
iv) At low initial temperatures, energy absorption is dominated by the fast-heating subsystem, leading to quicker and higher prethermalization.
}
\label{fig:model_outline}
\end{figure}

Here, a crucial point is the difference in the energy spectra of the two subsystems (see Fig.~\ref{fig:model_outline}):
a) the Hamiltonian $\hat{H}_{\mathrm{A}}$ on chain~A can access a broad range of energies, $[E^{\min}_{\mathrm{A}}, E^{\max}_{\mathrm{A}}]$.
b) the Hamiltonian $\hat{H}_{\mathrm{B}}$ on chain~B is limited to a narrower range, $[E^{\min}_{\mathrm{B}}, E^{\max}_{\mathrm{B}}]$, with $[E^{\min}_{\mathrm{A}}, E^{\max}_{\mathrm{A}}] \supset [E^{\min}_{\mathrm{B}}, E^{\max}_{\mathrm{B}}]$.
Moreover, both subsystems are subjected to periodic driving fields, say $\hat{V}_{\mathrm{A}}(t) + \hat{V}_{\mathrm{B}}(t)$, but with a key difference:
a) chain~A is driven with a frequency $\omega_{\mathrm{A}}$ such that it heats up rapidly.
b) chain~B is driven much more weakly (or not at all), with a much higher frequency $\omega_{\mathrm{B}} \gg \omega_{\mathrm{A}}$, so its heating is strongly suppressed~\cite{KUWAHARA201696,PhysRevLett.116.120401}.
These heating rates can be finely tuned by adjusting the drive frequencies.
Here, we have implicitly assumed that each subsystem is nonintegrable, so that at long times, Floquet eigenstate-thermalization hypothesis ensures true thermalization over the whole Hilbert space~\cite{PhysRevX.4.041048}.

The dynamics toward the prethermal states strongly depend on the initial particle distribution.
In a low-temperature (cold) state, most particles are in chain A ($N_{\mathrm{A}} \sim N$). Because chain A heats up rapidly, and its energy spectrum fully contains that of chain B ($[E^{\min}_{\mathrm{B}}, E^{\max}_{\mathrm{B}}] \subset [E^{\min}_{\mathrm{A}}, E^{\max}_{\mathrm{A}}]$), the system can absorb energy up to the broad upper band edge of chain A. 
In contrast, for a high-temperature (hot) initial state, particles are distributed more evenly between chains A and B.
Now, because chain B heats much more slowly, and some particles are initially ``stuck" in chain B, the overall relaxation toward the prethermal plateau is bottlenecked by the slow energy absorption in chain B. 
Consequently, both the speed and the extent of prethermal energy uptake are reduced compared to the cold initial state.
Thus, starting from a lower-energy state leads to both faster and higher energy uptake in the prethermal regime.

Such a mechanism can arise in a broad class of driven systems with a conserved quantity, including cold-atom ladders, superconducting qubit arrays with conserved particle number (or parity), and driven Rydberg chains, provided the mechanisms (i)-(iv) in Fig.~\ref{fig:model_outline} are realized.

\sectionprl{Summary}
We have introduced and established the concept of the prethermal IME, both by constructing an explicit toy model and by clarifying the general mechanism behind its emergence. Our findings extend the landscape of nonequilibrium thermalization phenomena and suggest a new direction for experimental studies, particularly in engineered quantum systems such as cold atoms.

An intriguing open question is whether an analogous (prethermal) Mpemba effect can occur in cooling processes.
Moreover, although in this work we have focused on prethermal IME as formulated by energy relaxation, the phenomenon can, in principle, be formulated more generally using mathematical notions such as distances in phase space or thermomajorization~\cite{VanVu2025PRL}. This would allow for a unified framework encompassing a wider range of observables and dynamical settings.
Finally, it would be of both conceptual and practical interest to explore potential applications of the prethermal Mpemba effect in the optimization of quantum control and quantum energy processing.
\\

\sectionprl{Acknowledgments}
This work was supported by the Japan Society for the Promotion of Science KAKENHI (Grant Nos.~JP23K03286, JP23K25796, and JP24K01329), the Japan Science and Technology Agency (JST) COI-NEXT (No.~JPMJPF2221), JST PRESTO (No. JPMJPR2116), and Exploratory Research for Advanced Technology (No. JPMJER2302). 
K. Sugimoto appreciates the support from the RIKEN TRIP initiative (RIKEN Quantum).
The numerical calculations were performed using the QuSpin library~\cite{quspin1, quspin2}.

\bibliography{paper.bbl}

\onecolumngrid
\vspace{10pt}

\begin{center}
{\large \bf End Matter}\\
\end{center}

\twocolumngrid

\sectionprl{Numerical details}
To calculate both imaginary- and real-time evolutions of canonical thermal pure quantum states, we employ the time-dependent Lanczos algorithm \cite{Park1986JCP}.
The time-step sizes are set to $\delta \tau = 0.05$ for imaginary-time evolution and $\delta t = T/100$ for real-time evolution, with the Krylov subspace dimension fixed at $m_{\text{L}}=10$.
Prior to time-evolution calculations, we prepare a random vector $\ket{\psi_0} = \sum_{\nu=1}^{D} c_{\nu} \ket{\nu}$, where $\left\{ \ket{\nu} \right\}$ denotes an orthonormal Fock basis spanning the Hilbert space whose dimension is $D = \binom{L}{N}$.
The complex coefficients $c_{\nu} = a_{\nu} + i b_{\nu}$ are constructed from real-valued random variables $a_{\nu}$ and $b_{\nu}$, independently generated from a Gaussian distribution with zero mean.

\begin{figure}[b]
\centering
\includegraphics[width=\columnwidth]{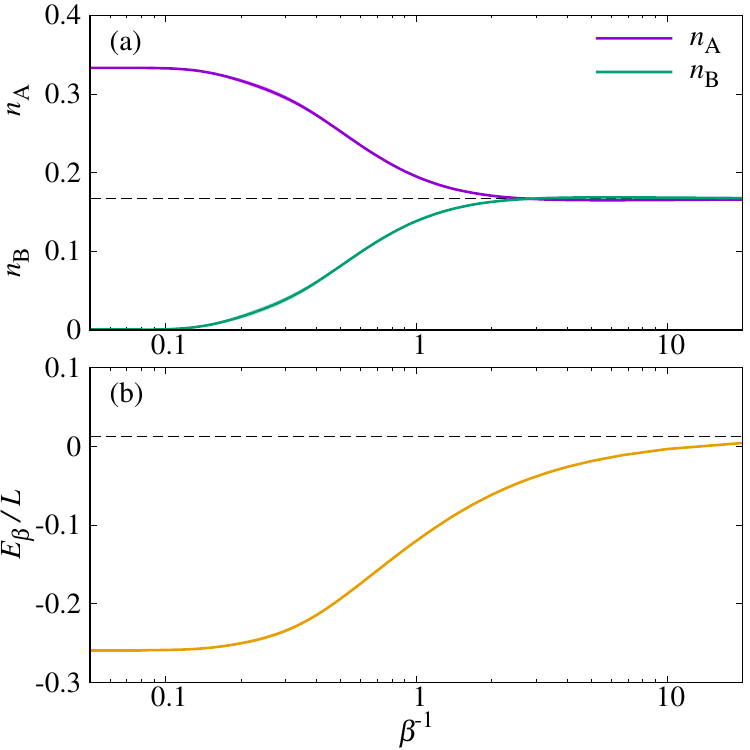}
\caption{
Temperature dependence of (a) the particle density per site and (b) the energy per site.
The dashed lines represent the corresponding values at infinite temperature.
}
\label{fig:equilibrium}
\end{figure}

Figure~\ref{fig:equilibrium}(a) shows the particle density per site in each chain,
$
  n_{\alpha} (\beta)
    = \frac{1}{L_{\alpha}} \sum_{j=1}^{L_\alpha} \Braket{\hat{n}_{j, \alpha}}_{\beta},
$
as a function of temperature $\beta^{-1}$.
Although the number of particles is equal in both chains at high temperatures, more particles accumulate in chain A as the temperature is lowered.
In particular, for $\beta^{-1} \lesssim 0.1$, almost all the particles are distributed in chain A.
Figure~\ref{fig:equilibrium}(b) shows the energy per site as a function of temperature.
It can be seen that the energy is almost equal to the ground-state energy for $\beta^{-1} \lesssim 0.1$.
\\

\sectionprl{Singular-value decomposition and normalized inverse-participation ratio for subsystems}
Since we consider an isolated quantum system, the total number of particles is conserved during time evolution.
That is, particle number conservation ensures that $\sum_{j} n_j = N$, where $n_j \in \left\{ 0,1 \right\}$ denotes the fermion occupation number at site $j$.
The quantum state of the system can thus be expressed as
\begin{equation}
  \ket{\psi}
    = \sum_{\substack{n_1, n_2, ..., n_L \\ \sum_j {n}_j = N}}
      C_{n_1, n_2, ..., n_L}
      \ket{n_1, n_2, ..., n_L},
\end{equation}
where $\ket{n_1, n_2, ..., n_L}$ denotes a number basis in the Fock space, and the normalized condition $\sum \left|C_{n_1, n_2, ..., n_L}\right|^2 = 1$ is satisfied.

\begin{figure}
  \centering
  \includegraphics[width=\columnwidth]{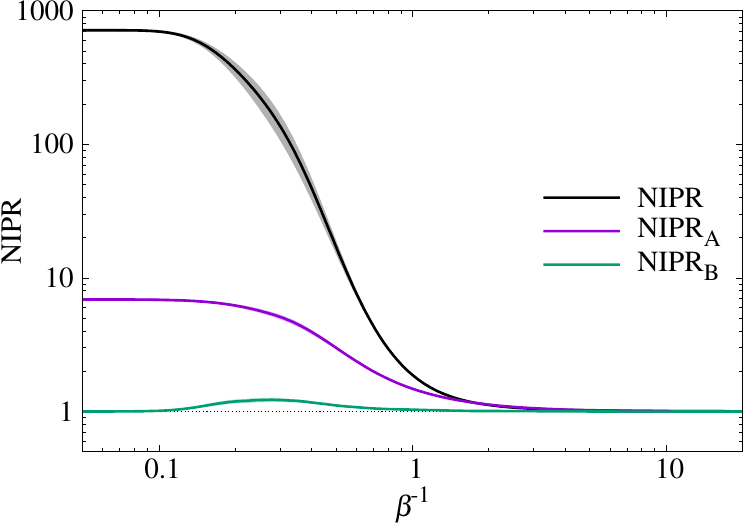}
  \caption{
  Temperature dependence of the NIPR.
  Chains A and B are considered subsystems.
  }
  \label{fig:nipr}
\end{figure}

We now divide the system into two subsystems, X and Y.
Notably, the number of particles within each subsystem is not fixed, although their sum is constrained by the total particle number.
Denoting the number of particles in subsystem X as $n$, we define composite indices for each subsystem: $\alpha_{n} = \left\{n_i \mid i \in \text{X} \right\}$ and $\beta_{n} = \left\{n_j \mid j \in \text{Y} \right\}$, with the constrains $\sum_{i \in X} n_i = n$ and $\sum_{j \in \text{Y}} n_{j} = N - n$.
By applying singular value decomposition, the coefficients can be expressed as
\begin{equation}
  C_{n_1, n_2, ..., n_L}
    = \sum_{n=\max(0, N - L_{\text{Y}})}^{\min(N, L_\text{X})} 
      \sum_{s_{n}} 
      U^{(n)}_{\alpha_{n}, s_{n}} \lambda^{(n)}_{s_{n}} \left(V^{(n)}_{\beta_{n}, s_{n}}\right)^*,
\end{equation}
where $L_{\text{X}}$ and $L_{\text{Y}}$ denotes the number of sites in subsystems X and Y, respectively.
This yields the Schmidt decomposition of the quantum state, given by
\begin{equation}
  \ket{\psi}
    = \sum_n \sum_{s_n}
      \lambda^{(n)}_{s_n}
        \ket{\phi^{(n)}_{s_n}}_{\text{X}}
        \ket{\varphi^{(n)}_{s_n}}_{\text{Y}},
\label{eq:oRBOvkMh}
\end{equation}
where the Schmidt basis states are defined as
\begin{align}
  \ket{\phi^{(n)}_{s_n}}_{\text{X}}
    &= \sum_{\alpha_n} U^{(n)}_{\alpha_{n}, s_{n}}
      \ket{\alpha_n}_{\text{X}}
\label{eq:klKt8RHT}
\\
  \ket{\varphi^{(n)}_{s_n}}_{\text{Y}}
    &= \sum_{\beta_n} \left(V^{(n)}_{\beta_{n}, s_{n}} \right)^*
      \ket{\beta_n}_{\text{Y}}.
\label{eq:jUXdIcUD}
\end{align}

In the main text, we introduce the normalized inverse participation ratio (NIPR) as a measure to quantify the thermalization properties of quantum states.
Based on the Schmidt decomposition given in Eqs.~\eqref{eq:klKt8RHT} and \eqref{eq:jUXdIcUD}, we define the NIPR for subsystems X and Y as 
\begin{align}
  \mathrm{NIPR}_{\text{X}}
    &= \sum_{n} 
      \sum_{s_{n}}
      \left(\lambda^{(n)}_{s_{n}} \right)^2
      \frac{D^{(n)}_{\text{X}} + 1}{2}
      \sum_{\alpha_{n}}
      \left|U^{(n)}_{\alpha_{n}, s_{n}}\right|^4
\\
  \mathrm{NIPR}_{\text{Y}}
    &= \sum_{n}
      \sum_{s_{n}}
      \left(\lambda^{(n)}_{s_{n}} \right)^2
      \frac{D^{(n)}_{\text{Y}} + 1}{2}
      \sum_{\beta_{n}}
      \left|V^{(n)}_{\beta_{n}, s_{n}}\right|^4,
\end{align}
where $D^{(n)}_{\text{X}} = \binom{L_{\rm X}}{n}$ and $D^{(n)}_{\text{Y}} = \binom{L_{\rm Y}}{N-n}$ are the dimensions of the Fock space of subsystems X and Y, respectively, for a given particle number $n$ in subsystem X.
Note that $D = \sum_n D^{(n)}_{\mathrm{X}} D^{(n)}_{\mathrm{Y}}$.

Figure~\ref{fig:nipr} shows the temperature dependence of the NIPR for the entire system and for the individual subsystems when partitioned into chains A and B.
In the high-temperature regime, all NIPR values are 1 since the quantum state is almost random.
As the temperature decreases, the NIPR for the entire system and for chain A exhibit a monotonic increase.
In contrast, the NIPR in chain B shows a nonmonotonic behavior: it peaks at intermediate temperatures, subsequently decreases, and eventually returns to $\mathrm{NIPR}_{\mathrm{B}} \approx 1$.
This trend arises from the fact that, as shown in Fig.~\ref{fig:equilibrium}(a), the state in chain B becomes uniquely determined to be the vacuum state with zero particles in the low-temperature limit.
Consequently, the NIPR serves as an effective diagnostic tool for assessing deviations from the infinite-temperature thermal state, particularly in the entire system and in chain A.
\\

\begin{figure}
  \centering
  \begin{overpic}[width=\columnwidth]{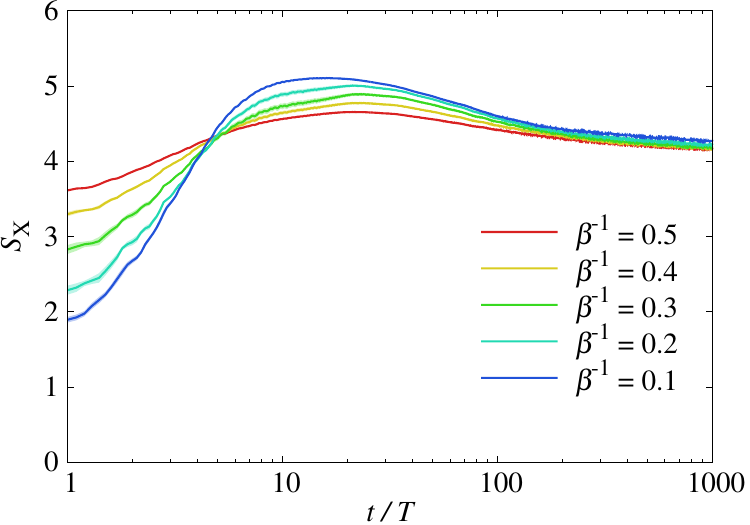}
    \put(18,18){\includegraphics[width=0.4\linewidth]{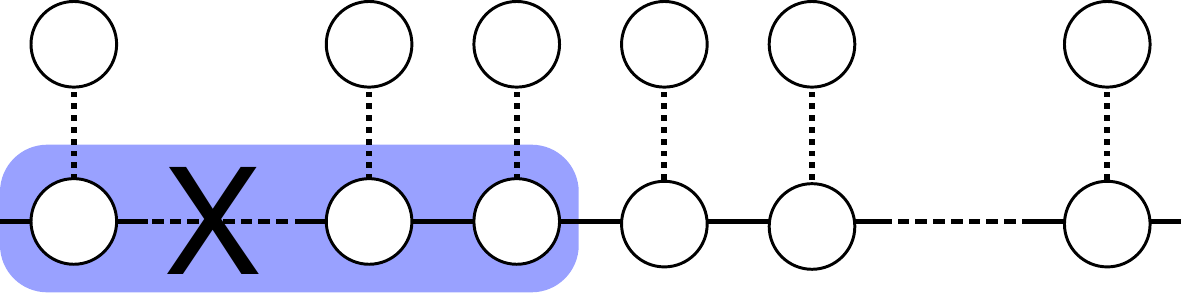}}
  \end{overpic}
  \caption{
  Temperature dependence of the entanglement entropy.
  The subsystem X corresponds to the left half of chain A, as illustrated in the inset.
  }
  \label{fig:dynamics_EE}
\end{figure}

\sectionprl{Inverse Mpemba effect in entanglement entropy}
The von Neumann entanglement entropy associated with the bipartition defined in Eq.~\eqref{eq:oRBOvkMh} is given by
\begin{equation}
  S_{\text{X}}
    = -\sum_{n} 
      \sum_{s_{n}}
      \left(\lambda^{(n)}_{s_{n}} \right)^2 \ln \left(\lambda^{(n)}_{s_{n}}\right)^2.
\end{equation}
To examine the entanglement within chain A, we choose the subsystem X as the left half of chain A.
Figure~\ref{fig:dynamics_EE} shows the time evolution of the entanglement entropy for several initial temperatures.
During the time evolution, the entanglement entropy for the initial temperature $\beta^{-1} = 0.1$ exhibits a pronounced increase and intersects with the entanglement entropies corresponding to higher initial temperatures.
This behavior indicates that, similar to the energy dynamics, the entanglement entropy also exhibits the prethermal IME.

\end{document}